\documentclass[fleqn,10pt]{wlscirep}
\usepackage[utf8]{inputenc}
\usepackage{siunitx}
\usepackage[T1]{fontenc}
\usepackage{enumitem}
\usepackage{lineno,hyperref}

\title{Detection of subtle cartilage and bone tissue degeneration in the equine joint using polarisation-sensitive optical coherence tomography}

\author[1,2,*]{Matthew Goodwin}
\author[3,a]{Marie Klufts}
\author[2,b]{Joshua Workman}
\author[2,c]{Ashvin Thambyah}
\author[1,d]{Fr\'{e}d\'{e}rique Vanholsbeeck}
\affil[1]{The Dodd-Walls Centre for Photonic and Quantum Technologies, Department of Physics, University of Auckland, Auckland 1010, New Zealand}
\affil[2]{Department of Chemical and Materials Engineering, University of Auckland, Auckland 1010, New Zealand}
\affil[3]{Institute of Biomedical Optics, University of L\"ubeck, 23562 L\"ubeck, Germany}

\keywords{Detection of subtle cartilage and bone tissue degeneration in the equine joint using polarisation-sensitive optical coherence tomography}

\begin{abstract}
\textbf{Objective:} To explore the ability of polarisation-sensitive optical coherence tomography (PS-OCT) to rapidly identify subtle signs of tissue degeneration in the equine joint.

\noindent \textbf{Design:} Polarisation-sensitive optical coherence tomography (PS-OCT) images were systematically acquired in four locations along the medial and lateral condyles of the third metacarpal bone in 5 equine specimens. Intensity and retardation PS-OCT images, and anomalies observed therein, were then compared and validated with high resolution images of the tissue sections obtained using Differential Interference contrast (DIC) optical light microscopy. 

\noindent \textbf{Results:} The PS-OCT system was capable of imaging the entire equine osteochondral unit, and allowed delineation of the three structurally differentiated zones of the joint, that is, the articular cartilage  matrix, zone of calcified cartilage and underlying subchondral bone. Importantly, PS-OCT imaging was able to detect underlying matrix and bone changes not visible without dissection and/or microscopy.

\noindent \textbf{Conclusion:} PS-OCT has substantial potential to detect, non-invasively, sub-surface microstructural changes that are known to be associated with the early stages of joint tissue degeneration.\\

\end{abstract}
\begin{document}
\flushbottom
\maketitle
\vspace{-15mm}
\section*{Introduction}

Osteoarthritis (OA) is a degenerative joint disease where gradual loss of the cartilage layer in articulating joints leads to severe pain and immobility. The relatively slow development of OA makes it commonly considered a disease of age and is often idiopathic. Once clinical symptoms arise, patients already exhibit substantial and irreversible joint degeneration. No cure exists, and the current invasive treatment options comes with significant socioeconomic costs~\cite{hunter2014individual}.
 
With OA research spanning well over one hundred years,\cite{brackett1915operative} the lack of a cure and limited treatment options reflect both the inherent complexity of the disease and our current limited understanding of the exact pathophysiology. Mechanical factors have long been considered one of the major sources of OA initiation, where overloading and acute injury have been linked with degeneration to the articular cartilage tissue~\cite{matthews1953composition, radin1971response}. By the turn of the century, several studies challenged the ’cartilage-centric’ notion of OA and the focus began to shift toward examining the underlying subchondral bone and its role as a potential OA precursor. Under this hypothesis, remodelling and subsequent stiffening of the subchondral bone hinders the shock-absorbing capacity of the entire joint. The reduced ability to dissipate energy during loading now results in physical damage to the overlying cartilage. This damage manifests in the form of cartilage fibrillation which has long been recognised as one of the first macroscopic signs of degeneration~\cite{radin1986role, bailey1997subchondral}.

The multi-faceted nature of OA makes it increasingly difficult to isolate the specific triggers of the disease. Response and adaption to mechanical stimuli occur in both the articular cartilage tissue and subchondral bone. Alterations to one component exacerbate problems with the other leading to a downward degenerative spiral~\cite{thambyah2007degeneration}. Recent research has highlighted the need to reexamine the disease from a system perspective rather than focus on the `bone-centric' or `cartilage-centric' theories~\cite{brandt2006yet, saxby2017osteoarthritis}. Lories and Luyten (2011)~\cite{lories2011bone} argue the entire definition of OA needs to evolve to incorporate both the cartilage and bone, and that future research needs to treat the disease in the context of a `bone-cartilage unit'.

Unravelling the complex interaction between cartilage and bone degeneration has proven to be challenging. The researcher needs to be able to identify the early structural changes occurring in the articular cartilage matrix, along with any alterations to the subchondral bone. Achieving this \textit{in vivo} compounds the problem by severely limiting the available tools. Arthroscopy is well-accepted as a standard for clinically assessing joint injury, yet it largely falls short of the researchers' requirements. Unfortunately, arthroscopy is restricted to simply examining the physical appearance within the joint and has a limited capacity to extract any meaningful information about the tissue integrity and functionality. Ultrasound or MRI have both been proven useful for clinical OA diagnosis. However, early-stage degenerative features occur on the micro- to nano-scale and are `invisible' to such modalities. The researcher is forced to use microscopy to explore the subtle changes occurring due to cartilage degeneration. Differential interference contrast (DIC) microscopy has increased in popularity recently due to its ability to study the tissue microstructure, unstained, and in a fully hydrated state~\cite{thambyah2006micro,thambyah2007degeneration,thambyah2009new}. DIC is highly sensitive to spatially changing refractive indices, and hence can image at high resolution, the structural detail in a tissue sample. By incorporating polarisation, details of the structural architecture can also be obtained. Specific to equine tissue, the present team has used DIC imaging technique to capture the entire full thickness of cartilage on bone at high resolution~\cite{turley2014microstructural}. Unfortunately, such microscopy protocols come at the cost of being destructive and also requiring significant time investment from trained users. What is required is a high-resolution, non-destructive imaging technique that is able to interrogate the tissue and identify degenerative changes even when the tissue appears macroscopically healthy.

Optical coherence tomography (OCT) has emerged as potential candidate to fill this imaging gap. The modality can be considered a light analogue of ultrasound; broadband light illuminates the sample and, depending on the optical properties, the backscattered light enables the reconstruction of an image. While conventional OCT measures the intensity of the reflected light to create a structural image, the polarisation of the reflected light can also be analysed to gain functional information. By irradiating the sample with light of a known polarisation, any change to the reflected light must be due to the sample and hence the birefringent properties of the tissue can be extracted. In the case of cartilage, the highly organised nature of the collagen network acts as the intrinsic contrast agent for polarisation sensitive OCT (PS-OCT). Overall, PS-OCT offers a route capable of providing insight on both the structure and integrity of the cartilage-bone unit as a whole.

The biggest limitation of OCT is its imaging depth. Near-infrared light is readily attenuated by biological tissue and therefore imaging is generally limited to less than two millimeters. Equine articular cartilage is comparably thinner than human and enables OCT to image the entire matrix along with upper portion of the subchondral bone plate. Furthermore, the abundant supply and presence of naturally occurring degeneration makes equine tissue a convenient model for exploratory studies.

Studying OA with (PS)-OCT has been done previously by several groups, most with the ultimate objective to characterise and quantify human cartilage degeneration~\cite{brill20153d,nebelung2015three,brill2015optical,ugryumova2005collagen,saarakkala2009quantification,nebelung2014morphometric,nebelung2016towards,nakamura2019changes,huang2011quantification,brill2016polarization,zhou2019slope,zhou2020detecting,goodwin2018quantifying}. While these studies present promising results, they largely focus on differentiating cartilage based on conventional OA grading schemes. These schemes often perform well when distinguishing between mild, moderate, and severe OA~\cite{ostergaard1999validity} but there is increasing interest in identifying the `pre-OA' state. te Moller et al. (2013)\cite{te2013arthroscopic} previously demonstrated the ability of OCT to be incorporated into an arthroscopic device to assess lesions in equine tissue. While they did identify subtle anomalies using OCT, no histology was was performed to validate the findings and was outside of the scope of their study. Further, no study has presented validated results of equine lesion assessment using both intensity and polarisation-sensitive OCT. Understanding the true potential of PS-OCT to detect the subtle changes that precede any categorical classification of OA is of great interest to the researcher and clinician.  

Therefore, the aim of this study is to evaluate the ability of PS-OCT to rapidly detect the subtle changes that are often overlooked. High resolution DIC microscopy was used to validate the findings and correlate the macro level health and any underlying pathology to the PS-OCT images.

\begin{figure}[ht]
\centering
\includegraphics[width=\linewidth]{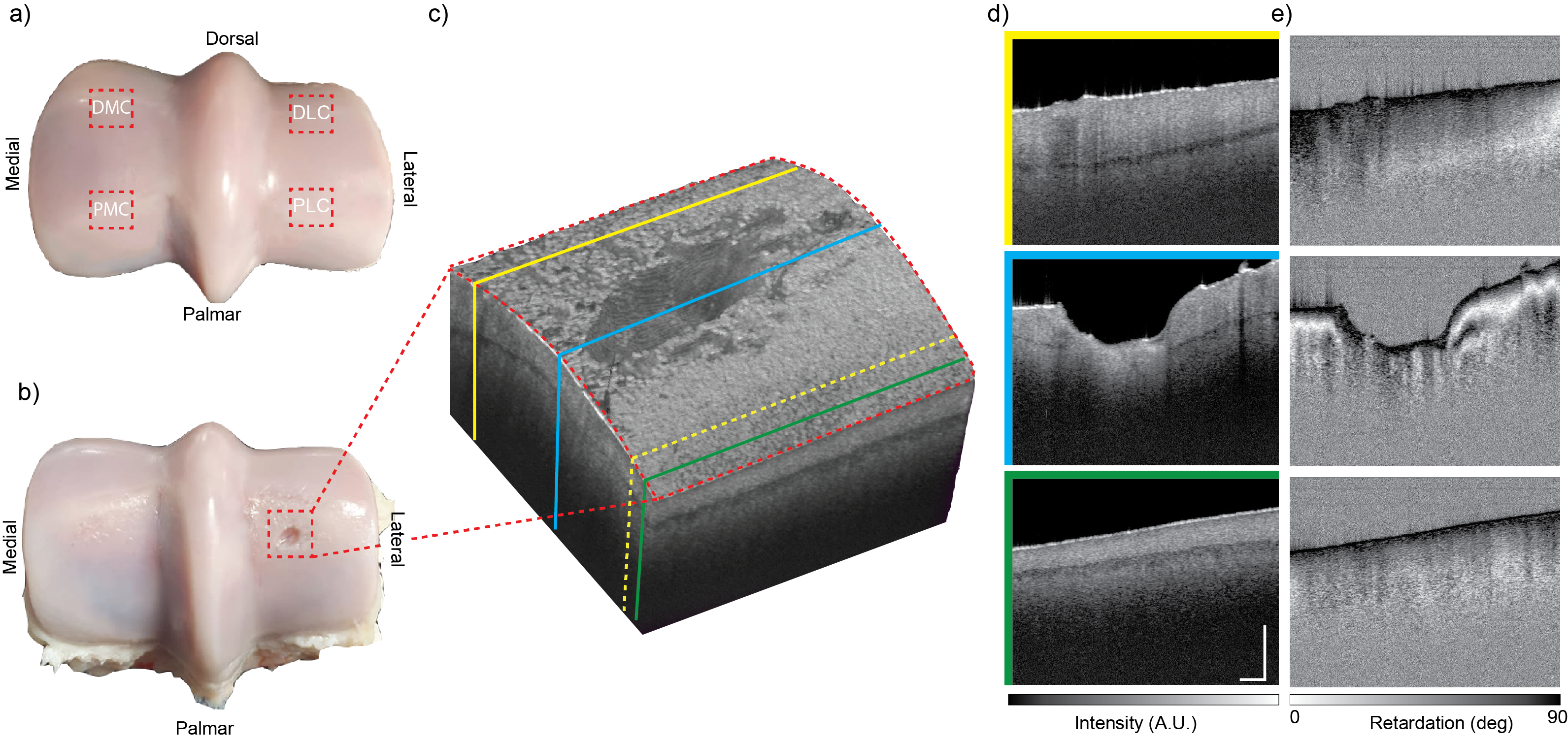}
\caption{(a) Macroscopic image of a healthy MC3 sample illustrating the four regions of interest. (b) Macroscopic image of MC3 sample with obvious cartilage damage on the palmar lateral condyle. (c) Three dimensional (C-scan), intensity OCT image corresponding to the dashed yellow box shown in (b).  Two dimensional (B-scan) intensity (d) and retardation images (e) taken from along multiple slices indicated on (c). Scale bars correspond to 1~mm. The dashed yellow line indicates the location of the images displayed in figure~\ref{fig:OCTSBRemodel}.}
\label{fig:OCTMethod}
\end{figure}
\section*{Materials and methods}
\subsection*{Specimens}

Five equine metacarpophalangeal joints, with undocumented histories, were obtained from a local slaughterhouse and stored at $-20^{\circ}$C. Each joint was thawed under running cold water before the third metacarpal bone (MC3) was dissected free. Any excess soft tissue and diaphyseal bone was removed. Two locations were chosen from each of the lateral and medial condyles, creating four regions of interest (ROI): palmar lateral condyle (PLC), palmar medial condyle (PMC), dorsal lateral condyle (DLC), and dorsal medial condyle (DMC) (Figure~\ref{fig:OCTMethod}a). Each region was grossly examined for signs of damage and degeneration. Following dissection, the MC3 joints were kept in 0.15~M saline while they were not being imaged. 

\subsection*{Optical coherence tomography imaging}

A custom PS-OCT system, described previously~\cite{thampi2020towards}, was used for the imaging. Briefly, the system uses a 1310~nm swept-source laser with a bandwidth of 100~nm and a 50~kHz sweep rate. The polarisation-sensitivity of the system allows intensity and retardation images to be acquired simultaneously. The axial and lateral resolution of the system is 12~$\mu$m and 40~$\mu$m respectively. The setup employs a 2-axis scanning galvanometer mirror system, allowing three dimensional images (C-scans) to be reconstructed and analysed. Total acquisition time for a volumetric image was less than 3 seconds. Similar to above, each OCT volume was qualitatively examined to identify anomalies.

The samples were taken out of the solution and excess water was removed to reduce saturation of the detectors. The four ROIs were imaged consecutively before the sample was returned to the saline solution in preparation for microscopy. The total time out of saline was less than 2 minutes.

%ImageJ prep Int: Min 70, Max 106

\subsection*{Microscopy}
To obtain microstructural data of the regions imaged with PS-OCT, all samples were prepared and sectioned for microscopic analysis using a DIC microscope (DIC, Nikon Eclipse 80i, Nikon Instruments). The resolution of microscopy is an order of magnitude better than PS-OCT, operating on the nanometer scale.

To prepare the sample for microscopy, each ROI was sawn (using a hacksaw) from the joint creating approximately 14~mm $\times$ 14~mm $\times$ 5~mm osteochondral blocks. Each block underwent chemical fixation using 10\% formalin for at least 48 hours and then decalcified in 10\% formic acid for 10 days which was sufficient time for the blocks to be easily sectioned. The samples were finally rinsed to neutralise any remaining acid. The region of interest was then cut from the osteochondral block using a scalpel and serially cryo-sectioned using a sledging microtome. The 30~$\mathrm{\mu}$m thick full-depth sections were wet mounted in 0.15 M saline solution, allowing them to be examined in a fully hydrated state. Image maps were created by stitching individual micrographs using image processing software (Adobe Photoshop 2020, Adobe Systems Incorporated, San Jose, California, US).

\section*{Results}
\begin{figure}[ht]
\centering
\includegraphics[width=\linewidth]{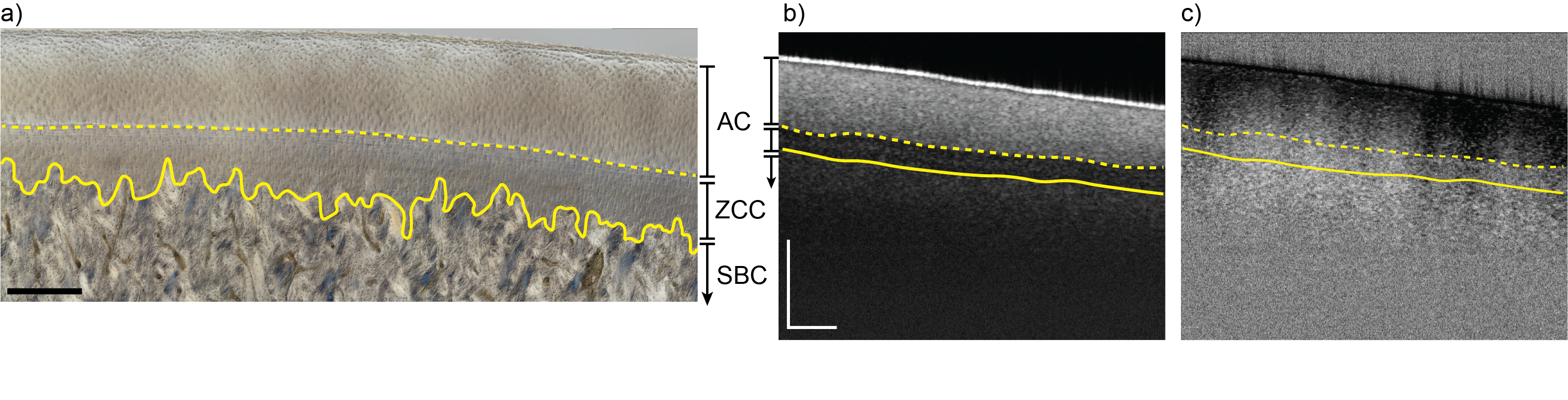}
\caption{An example of a state I sample. (a) DIC image of a healthy osteochondral unit comprised of articular cartilage (AC), zone of calcified cartilage (ZCC) and subchondral bone (SBC). The tidemark and the cement line are the interfaces between the sections and have been highlighted with a dashed and solid line respectively (scale bar = 0.5~mm). (b) Structural OCT images allow the three zones (AC, ZCC, SBC) to be segmented (scale bars = 1~mm) . (c) Retardation images of healthy cartilage exhibits little birefringence.}
\label{fig:OCTHealthy}
\end{figure}

A total of 20 regions of interest were evaluated using PS-OCT from the 5 equine samples. For any particular region imaged, based on the gross appearance of the tissue and the corresponding OCT images, three unique states emerged that were confirmed using DIC. The three states are:

\begin{enumerate}[label=\Roman*]
    \item Healthy appearing, determined via both gross examination and OCT.
    \item Healthy appearing via gross examination, but unhealthy appearing via OCT.
    \item Unhealthy appearing via both gross examination and OCT.
\end{enumerate}

Healthy appearing in the context of gross-examination refers to the surface of the cartilage appearing healthy and intact. Whereas, healthy appearing in the context of OCT refers to the entire cross-sectional intensity and retardation images appearing healthy. Tissue classified as state I, healthy appearing in both modes, corresponds to tissue that is intact and acts as the ground-truth `healthy' state. 

Any anomalies seen via gross examination or OCT leads to the classification being either state II or III. A state II classification is made when the tissue appears healthy via gross examination, but abnormalities can be seen in either (or both) the intensity and retardation images. Thus state II classification represents tissue that would likely be misinterpreted as being healthy via macroscopic observation techniques. Tissue that appears to be unhealthy in both modes is classed as state III and corresponds to tissue that has visible signs of degeneration through both gross examination and OCT.

\subsection*{State I. Healthy via gross examination and OCT}

On the microscopic level, a healthy osteochondral unit can be visually segregated into three primary regions: the articular cartilage (AC), the zone of calcified cartilage (ZCC), and the underlying subchondral bone (SBC) (Figure~\ref{fig:OCTHealthy}a). DIC imaging allows visualisation of all three regions along with the interface between each: the tidemark and cement line respectively. The articular cartilage surface appears smooth and the lines of chondrocyte continuity allow the traditional Benninghoff arcade~\cite{benninghoff1925form} to be observed. The calcification associated with the ZCC provides contrast against the AC, enabling identification of the tidemark which largely runs parallel to the articular surface morphology. The interface between the ZCC and bone, known as the cement line, is considerably more undulating compared to the smooth tidemark. 

In structural OCT images of healthy tissue (Figure~\ref{fig:OCTHealthy}b), the articular cartilage surface, tidemark, and cement line appear as continuous interfaces. The articular surface generally has the highest signal due to the large reflectivity coefficient as the light transitions from air to tissue. The matrix appears as a relatively homogeneous, highly scattering media whereas the ZCC and SBC scatter less. Reliably imaging the cement line and underlying subchondral bone may be challenging in certain circumstances due to a  range of factors affecting the light attenuation (e.g. topographical variations in cartilage thickness, relative beam-surface angle, etc.), however, it can generally be qualitatively discerned. Polarisation-sensitive images of healthy cartilage exhibit little birefringence (little change in the retardation) due to the optical axis of the tissue primarily being aligned with the deep zone collagen fiber orientation (Figure~\ref{fig:OCTHealthy}c) and therefore aligned with the optical beam. Overall this sample was validated as state I.

\begin{figure}[htb]
\centering
\includegraphics[width=\linewidth]{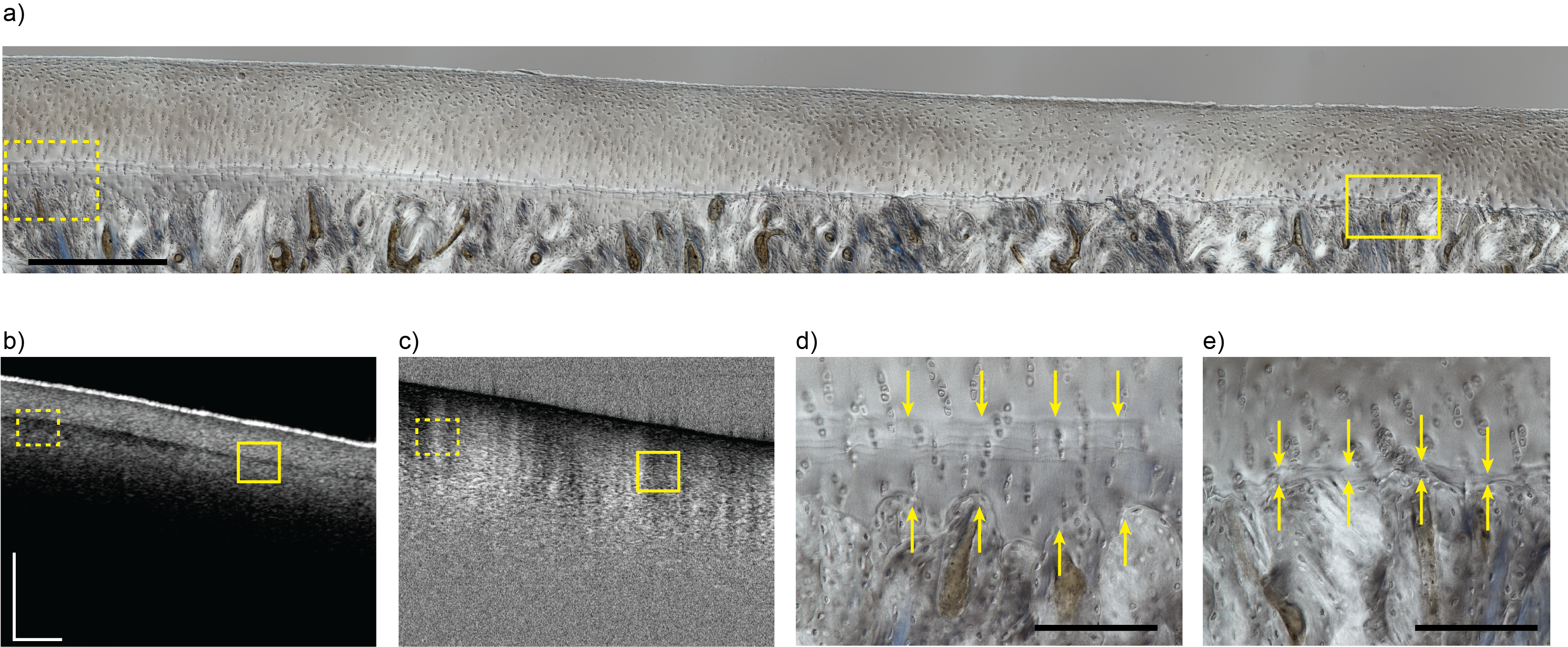}
\caption{An example of a state II sample. Images obtained from the section plane indicated by the yellow-dashed line shown in Figure~\ref{fig:OCTMethod}b, adjacent to a lesion. (a) DIC image showing localised subchondral differences (scale bar = 1~mm).  Compare the ZCC region shown in the dashed yellow box (d), versus that in the solid yellow box (e) (scale bar = 0.2~mm). Here the yellow arrows highlight the boundaries of the cement line and tidemark.  (b) Corresponding intensity OCT image highlights the variation in ZCC thickness (scale bars = 1~mm). (c) retardation image shows no remarkable changes in the articular cartilage matrix.}
\label{fig:OCTSBRemodel}
\end{figure}

\subsection*{State II. Healthy via gross examination, unhealthy via OCT}

There were several instances where regions appeared healthy via gross examination, however, subsequent OCT imaging revealed abnormalities that ranged from subtle changes affecting a single zone or even intense changes that affected multiple zones across the bone-cartilage unit.

The images shown in figure~\ref{fig:OCTSBRemodel} were taken from a palmar-lateral condyle adjacent to a large focal defect (refer to  figure~\ref{fig:OCTMethod}b). The DIC and (PS)-OCT images correspond to the 'healthy-appearing' region several millimeters away from the lesion and is marked with a dashed yellow line (figure~\ref{fig:OCTMethod}c). This sub-region showed no signs of surface fibrillation and both the DIC and OCT images indicate a pristine articular surface. The extracellular matrix appears intact and is in good agreement with the retardation image exhibiting little birefringence. The ZCC has lower scattering compared to the adjacent AC and SBC. On the left side of the intensity OCT image, the ZCC appears of normal thickness and morphology, however it drastically narrows toward the medial side where the ZCC is almost indistinguishable, and the subchondral bone is adjacent to the articular cartilage matrix. Due to the sub-surface abnormality, this sample was classified as state II and later confirmed by the DIC images (figure~\ref{fig:OCTSBRemodel}d, and e).

\begin{figure}[ht]
\centering
\includegraphics[width=\linewidth]{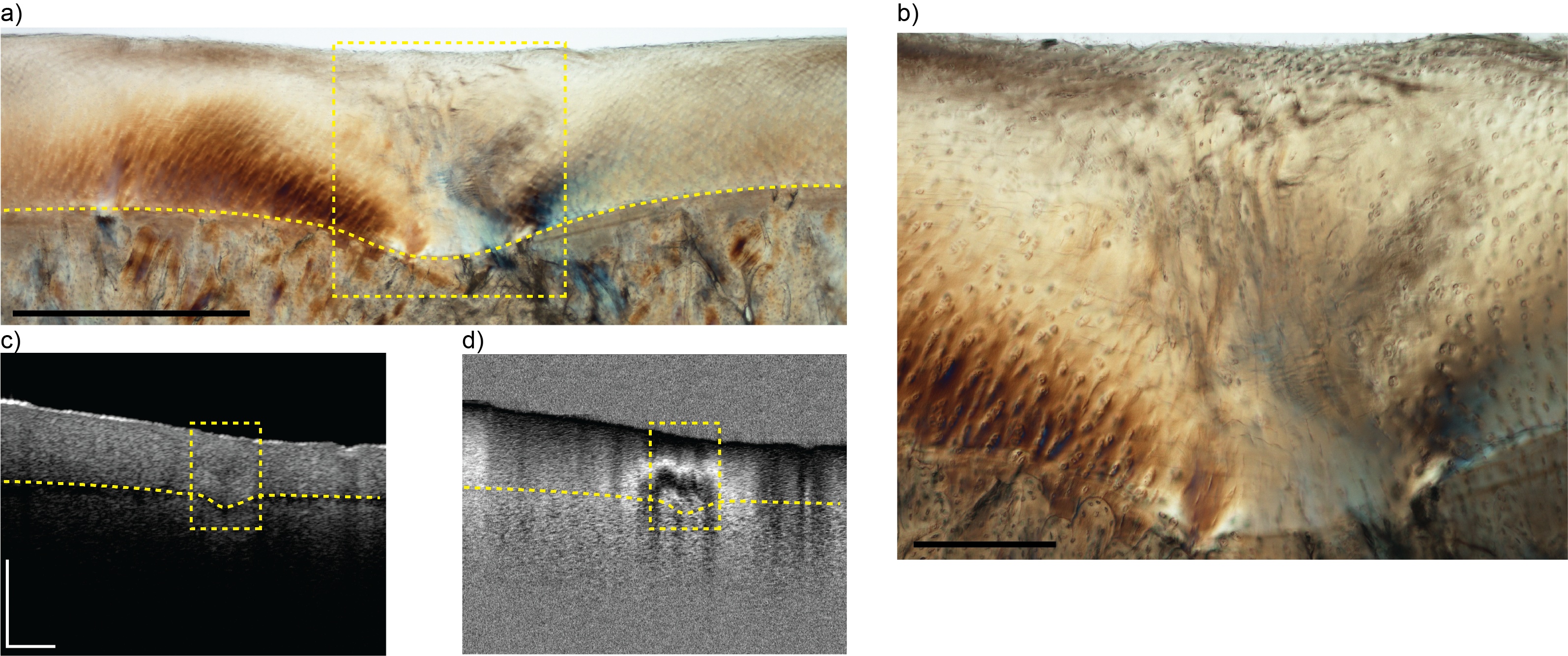}
\caption{An example of a state II sample. (a) DIC image reveals damaged cartilage matrix and appearance of a compression subchondral bone region (scale bar= 1~mm). (b) Higher magnification image of boxed region in (a) shows substantial matrix texture and abnormal chondrocyte orientation (scale bar = 0.25~mm). (c) Corresponding OCT image detects the changes to the cartilage-bone interface while the overlying cartilage appears normal (scale bar = 1~mm). (d) Corresponding retardation image shows significant birefringence is observable in the localised disruption region. The boxed regions in the two OCT images correspond to that in (a).}
\label{fig:OCTdisruptHealthy}
\end{figure}

A second state II sample, shown in figure~\ref{fig:OCTdisruptHealthy}, was taken from the palmar-lateral condyle. However this sample displayed  only very mild, localised surface fibrillation. The images were taken several millimeters away from the surface fibrillation, from a region that appeared healthy. Both OCT and DIC techniques revealed an intact surface with substantial focal disruption to both the underlying matrix and subchondral bone. The intensity OCT image (figure~\ref{fig:OCTdisruptHealthy}c) alone provides little insight into the collagen network integrity, exhibiting a typical matrix scattering pattern. However, the intensity image does detect localised compression of the cartilage-bone interface. The retardation images (figure~\ref{fig:OCTdisruptHealthy}d) presents strong birefringence, as characterised by the observable banding pattern. However, the high birefringence remains localised with the AC matrix on either side of the indentation exhibiting minimal optical activity, suggesting the surrounding matrix is relatively intact. The DIC image confirm these findings and reveals extensive localised disorganisation, displaying both a textured matrix and decreased chondrocyte density in the central area and then returns to normal appearance in the adjacent regions.

%Peak AC thickness is 852 micron

\begin{figure}[htb]
\centering
\includegraphics[width=\linewidth]{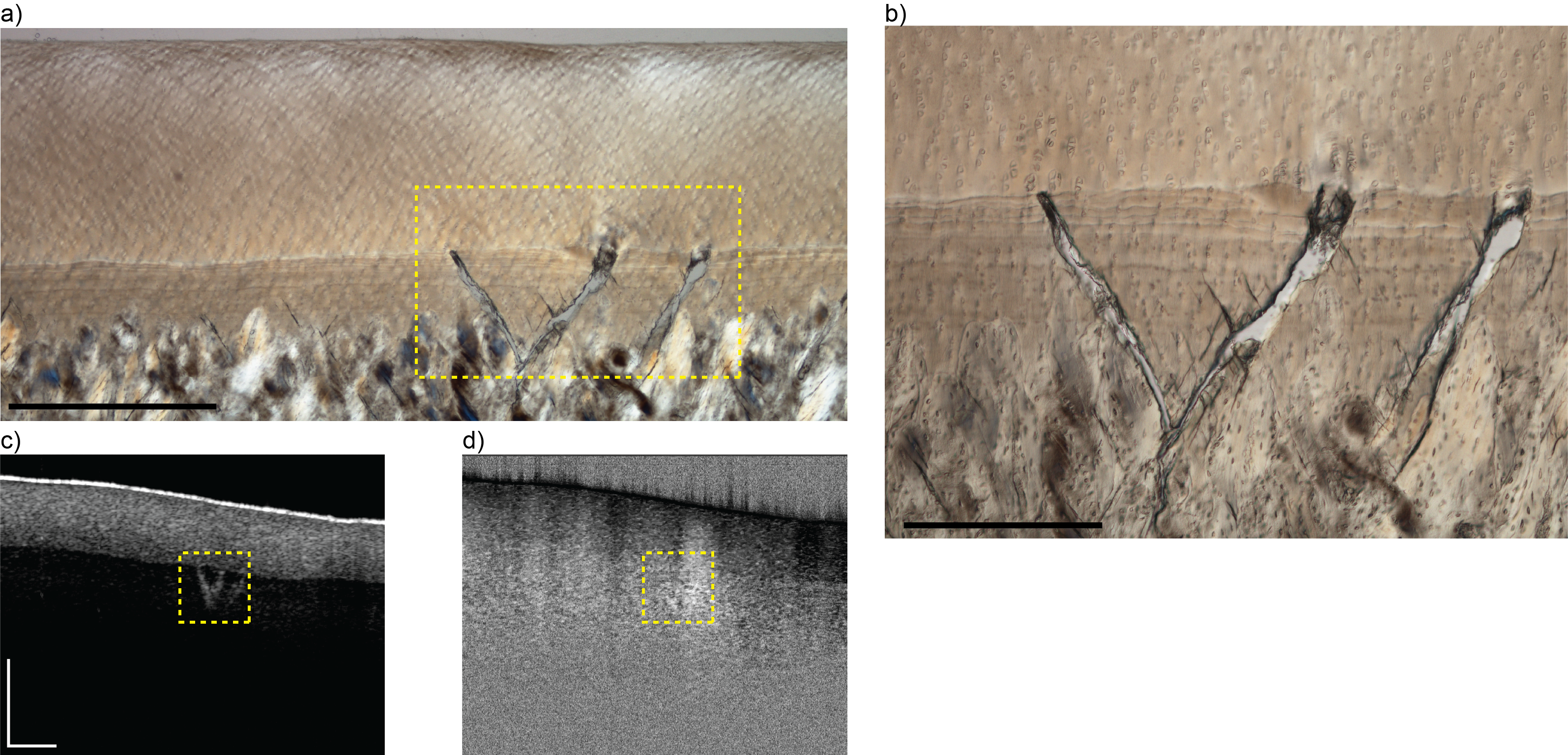}
\caption{An example of a state II sample. a) DIC image of healthy-appearing articular cartilage with multiple long cavities observable extending from the subchondral bone (scale bar = 0.5~mm). b) Higher magnification DIC image of the boxed section shown in (a) (scale bar  = 0.250~mm). The corresponding intensity OCT image (c) displays two hyper-intense regions that have the same morphology as the cavities seen in the DIC images. b) The retardation image provides little extra ability to identify the anomalies (scale bar = 1~mm). The boxed regions in the two OCT images correspond to that in (a).}
\label{fig:OCT6Crack}
\end{figure}

Figure~\ref{fig:OCT6Crack} presents a sample from the palmar-lateral condyle that appears macroscopically healthy. DIC microscopy indicates the articular cartilage layer is in pristine condition; both the surface and the matrix are perfectly intact. The intensity and retardation images support this claim with a typical matrix scattering pattern and little observable birefringence. Interestingly, two hyper-intense anomalies are visible in the intensity OCT images in the region beneath the tidemark. Confirmed by DIC microscopy, the features correspond to long linear cavities that originate from the subchondral bone and breach the uppermost tidemark or the boundary between the zone of calcified cartilage and hyaline cartilage proper. These linear cavities are noticeably sitting above a bed of microcracks in the subchondral bone. The significance of these unique features and the ability to rapidly identify them is discussed later in this paper.

%442 pixels between is 368 microns between V

\subsection*{State III. Unhealthy via gross examination and OCT}
More often, large anomalies observable via OCT are also observable on a macroscopic scale via gross examination. However, it is often difficult to understand the extent and significance of the damage.

\begin{figure}[htb]
\centering
\includegraphics[width=\linewidth]{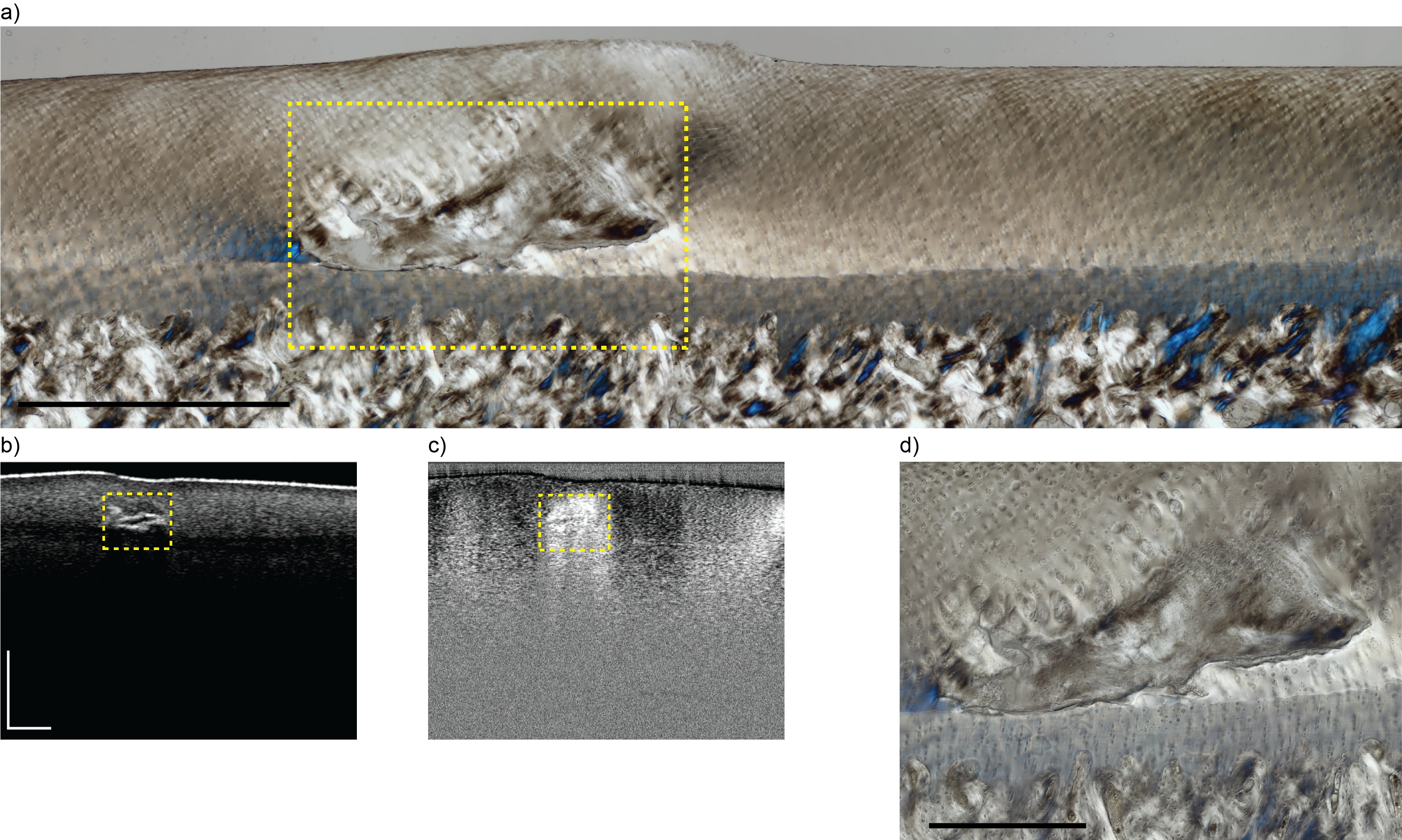}
\caption{An example of a state III sample. (a) DIC image of an acute sub-surface cartilage injury (scale bar = 1~mm). (b) Such damage is clearly evident via OCT imaging (scale bar = 1~mm). (c) Corresponding retardation image exhibits mild optical activity around the site of the lesion. (d) Higher magnification DIC image of boxed region in (a) shows chondrocyte cloning and attempted repair around the focal defect (scale bar = 0.5~mm). The boxed regions in the two OCT images correspond to that in (a).}
\label{fig:OCTSevereSubsurface}
\end{figure}

What appears to be some ‘healing’ following an injury in the cartilage matrix is shown in Figure~\ref{fig:OCTSevereSubsurface}. Here, a sub-surface focal defect is present deep in the articular cartilage matrix on the palmar-medial condyle. The slight irregularity of the surface profile means this damage is macroscopically visible and classed as state III. However the degree and severity of the defect was unknown. The intensity OCT image (figure~\ref{fig:OCTSevereSubsurface}b) provides clear information on the morphology of the defect while the moderate localised birefringence in the retardation  images (figure~\ref{fig:OCTSevereSubsurface}c) indicates the integrity of the above tissue may also be compromised. The low birefringence in the adjacent matrix suggests the matrix is intact, in agreement with the DIC image. The higher magnification DIC image highlights that the defect does not penetrate into the ZCC and that the matrix immediately surrounding the lesion exhibits chondrocyte multiplication.

%1.3mm edge to edge defect

\begin{figure}[htb]
\centering
\includegraphics[width=\linewidth]{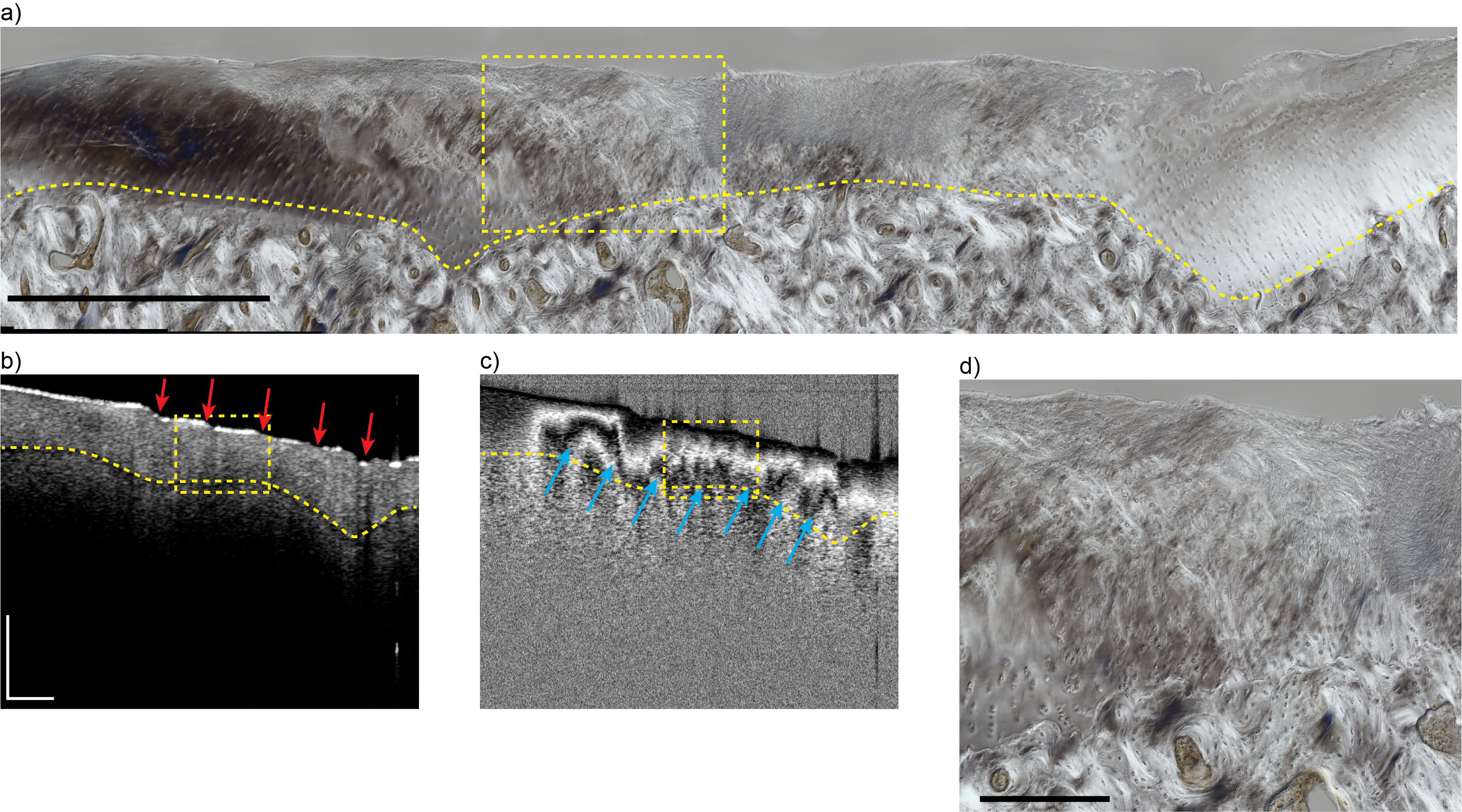}
\caption{An example of a state III sample. (a) DIC image shows signs of surface fibrillation, disorganised matrix, attempted repair, and subchondral remodelling (scale bar = 1~mm). (b) Corresponding intensity OCT image shows consistent trends in regards to the surface irregularity (red arrows) and cartilage-bone interface disruption (scale bar = 1~mm). (c) Corresponding retardation  image highlights the severeness of the disrupted matrix (blue arrows). d) Higher magnification DIC image of the boxed region shown in (a) emphasizes the highly textured matrix and fibrous repair tissue (scale bar = 0.250~mm).}
\label{fig:OCTfibrillationandDisruption}
\end{figure}

The sample displayed in Figure~\ref{fig:OCTfibrillationandDisruption} presented macroscopically with substantial surface irregularity in the palmar-lateral condyle (state III). The intensity OCT image was consistent with the macroscopic finding, but it also revealed there was severe disruption to the cartilage-bone interface. The tissue was strongly birefringent as indicated by the banding pattern in the retardation  image, suggesting extensive matrix disorganisation. DIC microscopy confirmed there was significant sub-surface damage and complete cartilage disruption. The affected matrix region shows strong matrix fibrosity.

%Crater to crater is around 3mm
%Thickness from center of double crater is around 0.460mm

\begin{figure}[htb]
\centering
\includegraphics[width=\linewidth]{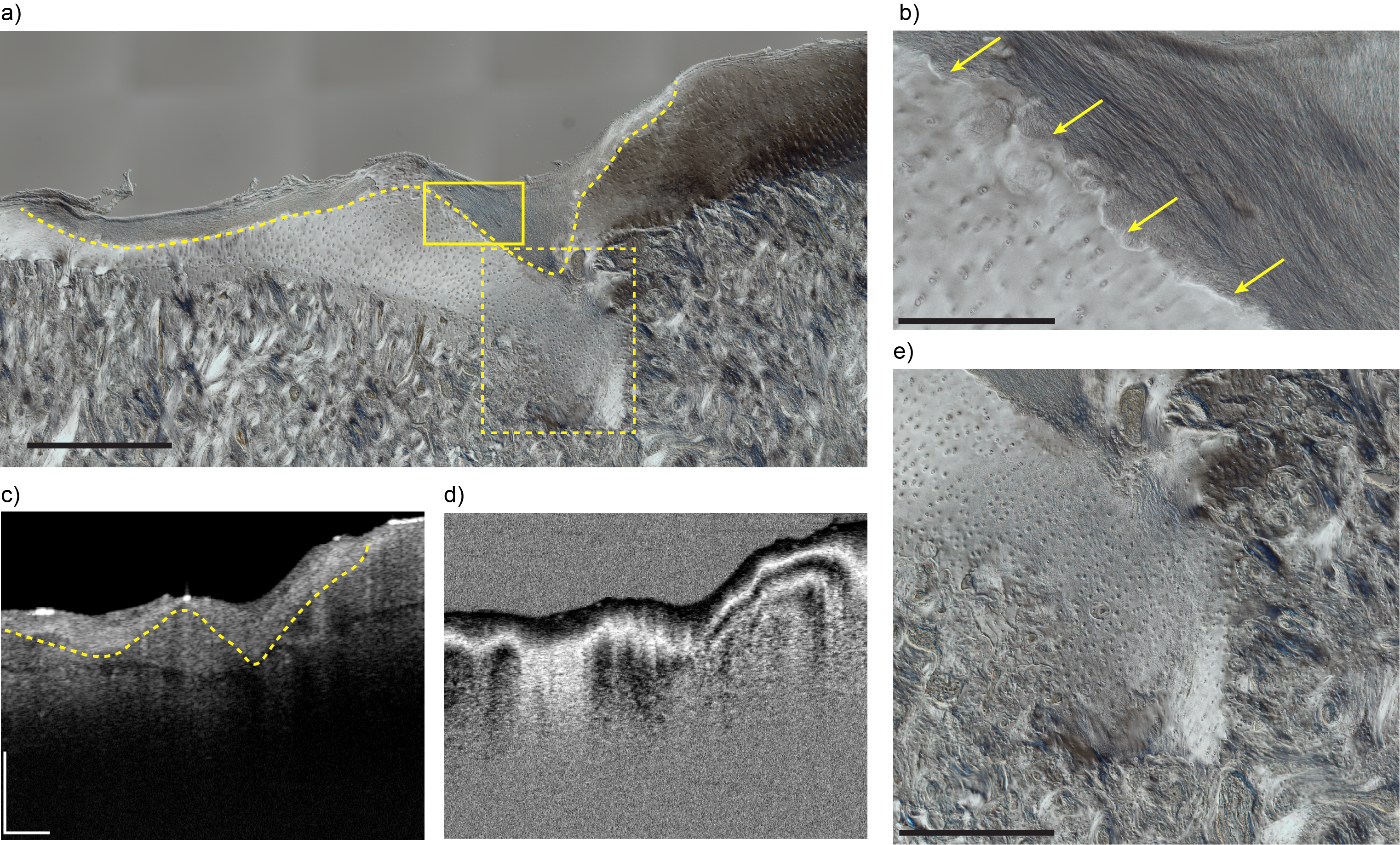}
\caption{An example of a state III sample. a) DIC image of articular cartilage showing characteristic signs of attempted repair (scale bar = 1~mm). b) Higher magnification DIC image of the solid box region shown in (a) highlights the structural discontinuity between the fibrous and native tissue (yellow arrow) (scale bar = 0.250~mm). c) OCT images show good agreement with the histological images however OCT fails to fully capture the cartilage-bone interface particularly in the region where the bone has collapsed (scale bar = 1~mm). (d) The banding patterns present in the polarisation-sensitive image is consistent with the matrix disruption. (e) Enlarged view of the solid boxed region in (a) showing subchondral bone collapse (scale bar = 0.5~mm).}
\label{fig:OCTfibrillationInhomogenieties}
\end{figure}

Finally of interest is a sample that displayed macroscopic signs of surface irregularity, however, the surface still appeared continuous and no fissures/full-thickness defects were observable (figure~\ref{fig:OCTfibrillationInhomogenieties}). Inhomogeneities within the scattering pattern of the intensity OCT image are visible along the width of the image (figure~\ref{fig:OCTfibrillationInhomogenieties}c). DIC microscopy revealed evidence of structural discontinuity between the fibrous repair tissue and the adjacent native hyaline cartilage. The morphology of the fibrous, possibly `neo-cartilage' is consistent with the OCT imhomogeneities and suggests differentiation between such repair tissue and native hyaline cartilage may be possible. The severe disruption to continuity of the subchondral bone results in the geometric distance from the cartilage surface to the cement line being greater than the imaging capabilities of OCT and thus the bone-cartilage interface can not be fully observed. The strong birefringence observed in the retardation images is consistent with the histological presentation of altered collagen organisation.

%is actually sample 104. From center of V shape fibers to far edge plateu of other fibers is 3.3mm. Plot E is 1mm from bean defect to bone. 0.36mm from edge of fiber dip to little part in fibers

\section*{Discussion}

% Direct traumatic injury to cartilage leads to a cascading degenerative spiral that includes chondrocyte death, subchondral microcracks, and disruption to the collagen matrix~\cite{lotz2010new}. The inherent complexity and variability in the degeneration pathway means that at any single point in time, a damaged joint presents as a superposition of the deterioration and healing processes. From a clinical perspective, non-destructively identifying and understanding the exact mechanisms at play within the tissue allows for optimal treatment decisions to be made.  PS-OCT has been proposed as a possible tool to fill this problem, however OCT tomograms reflect the optical properties of the tissue and can be difficult for a non-expert to interpret. In this study a range of degenerative features were identified with OCT and validated with DIC to highlight the potential of OCT to unravel and assess the extent of traumatic damage.

The PS-OCT system was capable of imaging the entire equine osteochondral unit, and allowed delineation of the three structurally differentiated zones of the joint, that is, the articular cartilage  matrix, zone of calcified cartilage and underlying subchondral bone. Importantly, PS-OCT imaging was able to detect underlying matrix and bone changes not visible without dissection and/or microscopy. The constituents that make up the extracellular matrix (e.g. chondrocytes, collagen, proteoglycans, etc.) are well below the resolution of OCT and therefore structural OCT imaging is often limited to extracting the optical parameters of the matrix or examining the integrity of the surface rather than imaging the cellular components. While surface irregularity metrics provide a quantitative method to assess early signs of fibrillation, it does not necessarily provide more information alone compared to arthroscopy. The true value of OCT lies in its ability to explore the changes occurring beneath the articular surface, down to the subchondral bone. Since equine AC is thinner compared to both human and bovine tissue, the zone of calcified cartilage and subchondral bone can be observed allowing full cartilage thickness to be determined~\cite{puhakka2016optical}.

% The aetiology of OA is still not well understood, however, recent research suggests that OA should be considered a disease of the whole joint~\cite{lories2011bone}. Previous studies have identified that subchondral remodelling and alterations to the calcified zone are among the earliest events of the disease~\cite{ding2003changes}. 

Topographical variations in the morphology of the osteochondral unit (e.g. ZCC narrowing seen in Figure~\ref{fig:OCTSBRemodel}) can lead to reduced load distribution capabilities and eventual deterioration of the overlying articular cartilage~\cite{burr2012bone}. The narrowing of the ZCC in figure~\ref{fig:OCTSBRemodel} occurs several millimeters from a full-thickness focal defect and while that area appears macroscopically healthy, the bone-cartilage unit may have impaired mechanical performance and hence be vulnerable to future degeneration.  

% Focal cartilage defect treatment methods have risen in popularity in recent years, where the treatment method often becomes an optimisation problem trying to retain as much healthy native tissue as possible. Understanding the spatial extent of the damage is critical when determining the optimal treatment plan and arming clinicians with such information has the potential to improve patient outcomes. 

% Arthroscopy is the most comparable technique to OCT in a clinical sense as both cannot image articular cartilage tissue without penetrating into the joint space. However, arthroscopic procedures have extremely limited capacity to assess any subsurface alterations.

In all of the presented figures, the surface is continuous to various extents; while there are macroscopic signs of surface fibrillation, the degree and significance of the damage is not fully appreciated until OCT/DIC is performed. Figure~\ref{fig:OCTdisruptHealthy} depicts a sample that appears microscopically intact, yet the OCT images suggest disruption to the matrix and underlying bone. It is suspected that an acute physical event has led to the compressed-appearance of the cartilage-bone interface (figure~\ref{fig:OCTdisruptHealthy}a) and the subsequent adaption of the matrix has led to the disrupted collagen network. In figure~\ref{fig:OCTSevereSubsurface}, it is possible a traumatic injury has caused delamination of the AC matrix at the tidemark. In both of the above cases PS-OCT clearly depicts the localised nature of the damage with the intensity images providing information on the structural morphology of the damage while the polarisation-sensitive images identify regions with disruption to the collagen network. 

The subchondral linear cavities that were imaged by OCT, and later confirmed by DIC (Figure~\ref{fig:OCT6Crack}), is an exciting find and demonstrate the ability of OCT to reveal such changes in the osteochondral junction in otherwise pristine appearing cartilage. These cavities are likely to be marrow filled. The obvious suggestion was that these linear cavities were a precedent to a bone marrow lesion. Bone marrow lesions (BMLs), are an important clinical sign of early osteoarthritis, and is detected using MRI~\cite{felson2001association}. The relationship between BMLs and subchondral microdamage in human knee joints has been established by ex-vivo studies~\cite{muratovic2018bone}, and they found that there was a strong correlation between the presence of BMLs and the frequency of linear cracks. In studying the sequela of osteoarthritis (OA), given that the presence and severity of BMLs are corelated with the progression of the disease~\cite{tanamas2010bone}, the ability of OCT to detect the presence of subchondral bone cavities may be a useful capability in future studies of OA initiation and progression. The other prospective consequence of these linear cavities, especially in the context of the equine bone and joint system, is in the subsequent catastrophic fractures that happens in horses following the accumulation of microcracks~\cite{muir2008exercise}. Could these linear cavities be a sign of increased vulnerability of the joint to failure? The ability of OCT to detect such changes sets it apart from other non-invasive techniques and highlights its effectiveness at providing insight into subtle degenerative changes that are often difficult to identify and potentially missed during traditional assessment techniques.

Figures~\ref{fig:OCTfibrillationandDisruption} \&~\ref{fig:OCTfibrillationInhomogenieties} display abnormalities that are less localised than previous cases and believed to be reflective of the natural cartilage adaption process following a certain degree of degeneration. In both cases there is severe disruption to the bone-cartilage interface, yet the two modes of adaptation appear different. In the former case, the surface remains relatively continuous while the articular cartilage exhibits extensive disorganisation with altered cellular and collagen structure. In comparison, the latter figure shows cartilage with reduced surface continuity and substantial evidence of a fibrous replacement matrix that has failed to integrate with the surrounding articular cartilage matrix. Due to the undocumented nature of the equine specimens, it is impossible to know the exact series of events leading to this point but it does highlight the potential of OCT to differentiate such mechanisms of repair and offer an opportunity to gain real-time insight into these processes in future studies.

The last decade has seen a field-wide adoption of quantitative analytic procedures to make the diagnosis process more objective. The inherent variability in multifaceted diseases such as cartilage degeneration makes definitively quantifying diseases difficult. Surface irregularity has been a popular metric employed to identify early stage primary OA using OCT. Unfortunately, this metric merely quantifies the changes apparent during arthroscopy, and recently there has been a greater emphasis on identifying biomarkers occurring before surface fibrillation. Michalik et al. (2019)\cite{michalik2019quantitative} developed a processing pipeline to quantitatively assess artificially induced subsurface defects with similar morphology of the one shown in figure~\ref{fig:OCTSevereSubsurface}. Alone, the framework provides little extra value in the wider diagnostic scope. Similarly, extracting birefringence as a measure of matrix integrity has been performed in multiple studies with various success~\cite{goodwin2018quantifying,brill2016polarization}. Each quantitative metric effectively provides information on a single part of the complex osteoarthritis puzzle. A goal of this study was to highlight and validate the key features that need to be considered during analysis and help future studies determine the correct combination of metrics that can be incorporated into a comprehensive OCT-OA assessment model.

While OCT may be unable to act as a truly non-invasive imagine technique, several studies proven its potential as an non-destructive arthroscopic tool in various mammalian species including human,\cite{chu2004arthroscopic,chu2010clinical} equine,\cite{te2013arthroscopic,niemela2014application} porcine,\cite{pan2003hand}. The most important finding in this study is the ability of PS-OCT to rapidly detect and understand the significance of subtle anomalies in the PS-OCT images. Histology-based processes require substantial time invested from the user both during the preparation and microscopy and would benefit greatly from PS-OCT as a rapid assessment tool. The results from this study have important implications for equine cartilage degeneration research, however, the translation of the results remains in question. Human articular cartilage is considerably thicker than equine cartilage and therefore OCT is often unable to image the cartilage-bone interface. Therefore the translatable degenerative features are those associated with the surface and cartilage matrix. Recent studies have shown that optical clearing techniques~\cite{bykov2016imaging} may allow the interface to be interrogated, however, this adds another degree of complexity and may limit the clinical suitability. Furthermore, this study only examined a total of five (n=5) MC3 joints which contained a range of observable degenerative features. This study does not present a comprehensive examination of all possible equine joint injuries but rather focuses on highlighting the variability of joint degeneration and the importance of understanding the potential and limitations of OCT imaging during assessment of equine cartilage degeneration.

\section*{Author contributions statement}
Conception and design of experiment: M.G, J.W, A.T, F.V

\noindent Conducted experiment: M.G, M.K

\noindent Analysis and interpretation of results: M.G, M.K, J.W, A.T, F.V

\noindent Drafting article: M.G, J.W, A.T, F.V

\noindent Final approval: A.T, F.V

\section*{Additional information}

\subsection*{Conflict of Interest}
The authors declare no competing interest, financial or otherwise.

\subsection*{Funding Sources}
The authors would like to acknowledge funding from Marsden Fund and Royal Society of New Zealand (UoA1509) which made this research possible.

\newpage

\bibliography{Equine}

\end{document}